\documentclass[12pt,thmsa,a4paper]{article}
\usepackage{amssymb}

\usepackage{sw20lart}

\input{tcilatex}
\begin{document}

\title{Form factor decomposition of generalized parton distributions at leading
twist}
\author{Ph. H\"{a}gler \\
%EndAName
Department of Physics and Astronomy, Vrije Universiteit\\
De Boelelaan 1081\\
1081 HV Amsterdam\\
The Netherlands}
\maketitle

\begin{abstract}
We extend the counting of generalized form factors presented in PRD\textbf{63%
}(2000) by Ji and Lebed to the axial vector and the tensor operator at
twist-2 level. Following this, a parameterization of all higher moments in $x
$ of the tensor (helicity flip) operator is given in terms of generalized
form factors.
\end{abstract}

\section{Counting generalized form factors}

Generalized parton distributions still attract increasing interest among
theorists and experimentalists alike investigating the quark and gluon
structure of hadrons. For a complete description of the nucleon structure at
(leading) twist 2 level, full knowledge of the corresponding spin
independent (vector) GPDs $H(x,\xi ,t)$ and $E(x,\xi ,t)$ \cite{Ji:1996ek},
the spin dependent (axial vector) GPDs $\widetilde{H}(x,\xi ,t)$ and $%
\widetilde{E}(x,\xi ,t)$ \cite{Ji:1996ek} as well as the transversity
(tensor, helicity-flip) GPDs\footnote{%
since $H_{T}(x,\xi \rightarrow 0,t\rightarrow 0)=h_{1}(x)=\delta q(x),$ we
will denote $H_{T}$ also as \emph{generalized transversity}} $H_{T}(x,\xi ,t)
$,$E_{T}(x,\xi ,t),\widetilde{H}_{T}(x,\xi ,t)$ and $\widetilde{E}_{T}(x,\xi
,t)$ \cite{Diehl:2001pm,Hoodbhoy:1998vm} is necessary. In numerous cases,
however, not the GPDs themselves, but their Mellin moments in $x$ are
needed, see e.g. the recent calculations of moments of GPDs in lattice QCD 
\cite{lattice1,lattice2,lattice3}. General higher Mellin moments of (matrix
elements of) bilocal operators lead to towers of local operators, which in
turn are parameterized in terms of generalized form factors (GFFs). The
correct counting of the number of independent generalized form factors is
quite important as a cross check of these parameterizations, as can be seen
e.g. from the mistaken application of time reversal in the case of the
helicity flip GPDs, see Ref. \cite{Hoodbhoy:1998vm}, which lead initially to
a wrong number of GPDs but has since then been corrected in \cite
{Diehl:2001pm}. Concerning the counting we will follow here closely the idea
presented in Ref. \cite{Ji:2000id} where it has been explicitely worked out
for the vector operator. There it is suggested that instead of studying
off-forward matrix elements like 
\begin{equation}
\left\langle P^{\prime }\right| \bar{\psi}(0)\Gamma iD^{\{\mu _{1}}iD^{\mu
_{2}}\cdots iD^{\mu _{n}\}}\psi (0)\left| P\right\rangle =a_{\Gamma }^{\mu
_{1}\ldots \mu _{n}}A(t)+b_{\Gamma }^{\mu _{1}\ldots \mu _{n}}B(t)+\cdots 
\label{ME1}
\end{equation}
under parity and time reversal in order to figure out in particular the
number of independent generalized\ form factors $A,B,\ldots $, one switches
to the crossed channel and considers the matrixelement 
\begin{equation}
\left\langle P\bar{P}\right| \bar{\psi}(0)\Gamma iD^{\{\mu _{1}}iD^{\mu
_{2}}\cdots iD^{\mu _{n}\}}\psi (0)\left| 0\right\rangle 
\end{equation}
Here and in the following $D=\overleftrightarrow{D}=1/2(\overrightarrow{D}-%
\overleftarrow{D}),$ while $\{\}$ stands for symmetrization. This procedure
reduces the counting to a matching of the $J^{PC}$-quantum numbers of the
nucleon-antinucleon state $\left\langle P\bar{P}\right| $ and the state
given by $\bar{\psi}(0)\Gamma iD^{\{\mu _{1}}iD^{\mu _{2}}\cdots iD^{\mu
_{n}\}}\psi (0)\left| 0\right\rangle $. We will see below, however, that the
number of GFFs depends on the operator (i.e. $\Gamma $), and that the
statement made in the last sentence of \cite{Ji:2000id} that ''...the number
of form factors of a twist-2, spin-$n$ operator is $n+1$ for nucleon
states.'' is wrong or at least misleading.

After having done the counting for the tensor operator $\Gamma \widehat{=}%
i\sigma ^{\mu \nu }$ (see table (\ref{countt}) and below), we finally
construct the parametrization of the corresponding tower of local operators
in terms of the the generalized form factors $A_{Tni}(t),\widetilde{A}%
_{Tni}(t),B_{Tni}(t)$ and $\widetilde{B}_{Tni}(t)$, Eq. (\ref{parat}).

\subsection{Nucleon-antinucleon states}

To get a list of the $P\bar{P}$-states we can e.g. follow the standard
textbook discussion of the (para-,ortho-)positronium states. The allowed $%
\left\langle P\bar{P}\right| $-states ( $P=(-)^{L+1},C=(-)^{L+S}$ ) with $%
J=\left| L-S\right| ,\ldots ,L+S$ are 
\begin{eqnarray}
&&
\begin{tabular}{llllll}
$S=0$ &  &  &  &  &  \\ \hline
\multicolumn{1}{|l}{$J^{PC}$} & \multicolumn{1}{|l}{$0^{-+}$} & 
\multicolumn{1}{|l}{$1^{+-}$} & \multicolumn{1}{|l}{$2^{-+}$} & 
\multicolumn{1}{|l}{$3^{+-}$} & \multicolumn{1}{|l|}{$\cdots $} \\ \hline
\multicolumn{1}{|l}{$L$} & \multicolumn{1}{|l}{$0$} & \multicolumn{1}{|l}{$1$%
} & \multicolumn{1}{|l}{$2$} & \multicolumn{1}{|l}{$3$} & 
\multicolumn{1}{|l|}{$\cdots $} \\ \hline
\end{tabular}
\text{ }  \nonumber \\
&&
\begin{tabular}{lllllllll}
$S=1$ &  &  &  &  &  &  &  &  \\ \hline
\multicolumn{1}{|l}{$J^{PC}$} & \multicolumn{1}{|l}{$0^{++}$} & 
\multicolumn{1}{|l}{$1^{--}$} & \multicolumn{1}{|l}{$1^{++}$} & 
\multicolumn{1}{|l}{$2^{--}$} & \multicolumn{1}{|l}{$2^{++}$} & 
\multicolumn{1}{|l}{$3^{--}$} & \multicolumn{1}{|l}{$3^{++}$} & 
\multicolumn{1}{|l|}{$\cdots $} \\ \hline
\multicolumn{1}{|l}{$L$} & \multicolumn{1}{|l}{$1$} & \multicolumn{1}{|l}{$%
0,2$} & \multicolumn{1}{|l}{$1$} & \multicolumn{1}{|l}{$2$} & 
\multicolumn{1}{|l}{$1,3$} & \multicolumn{1}{|l}{$2,4$} & 
\multicolumn{1}{|l}{$3$} & \multicolumn{1}{|l|}{$\cdots $} \\ \hline
\end{tabular}
\label{PPbar1}
\end{eqnarray}
The in principle accessible $0^{--}$-state is forbidden.

The following discussion is based in parts on Ref. \cite{Geyer:1999uq} and
Ref. \cite{Weinberg:mt}. Representations of the Lorentz group are denoted by 
$\left( A,B\right) $. The spin $j$ runs from $j=\left| A-B\right| ,\ldots
,A+B$.

\subsection{Vector operator}

As a warm-up exercise we summarize in this section the main findings of Ref. 
\cite{Ji:2000id}. The operator $v^{\mu }=\bar{\psi}(0)\gamma ^{\mu }\psi (0)$
corresponds to the $\left( \frac{1}{2},\frac{1}{2}\right) $ representation,
with two possible values of $j$, $j=0,1$. Here, $j=0$ corresponds to the
time-component $v^{0}$, while $j=1$ corresponds to the spatial-components $%
v^{i=1\ldots 3}$. This gives the $J^{PC}$s $1^{--}$ and $0^{+-}$. The more
general tower of operators 
\begin{equation}
\bar{\psi}(0)\gamma ^{\{\mu }iD^{\mu _{1}}iD^{\mu _{2}}\cdots iD^{\mu
_{n}\}}\psi (0)
\end{equation}
corresponds to 
\begin{equation}
\left( \frac{n+1}{2},\frac{n+1}{2}\right) ,
\end{equation}
with $j=0,1,\ldots ,n+1$. Here and in the following the subtraction of
traces is implicit. The $C$-parity is definite (independent of $\mu ,\mu
_{1}\ldots $ being $0$ or spatial) and given by $C=(-)^{n+1}$. The different
values of $j$ correspond to the individual indices $\mu ,\mu _{1}\ldots $
being $0$ or spatial (e.g. the case in which all indices are spatial
corresponds to the maximal $j$), and the parity is therefore dependent on $j$
and given by $P=(-)^{j}$ $(\hat{P}D^{0}\hat{P}^{-1}=D^{0},\hat{P}D^{i}\hat{P}%
^{-1}=-D^{i})$. This results in the angular momentum decomposition 
\begin{equation}
J^{PC}=j^{(-)^{j}(-)^{n+1}}=0^{(+)(-)^{n+1}},1^{(-)(-)^{n}}\ldots
,(n+1)^{(-)^{n+1}(-)^{n+1}}.
\end{equation}
A matching with the possible $\left\langle P\bar{P}\right| $-states gives
the number of available ''channels'', which is equal to the number of
generalized form factors 
\begin{equation}
\begin{tabular}{|l|l|l|l|l|l|l|c|}
\hline
$n\backslash J$ & $0$ & $1$ & $2$ & $3$ & $4$ & $\cdots $ & $\#$ of GFFs \\ 
\hline
$0$ & $0^{+-}$ & $1_{0,2}^{--}$ &  &  &  &  & $2$ \\ \hline
$1$ & $0_{1}^{++}$ & $1^{-+}$ & $2_{1,3}^{++}$ &  &  &  & $3$ \\ \hline
$2$ & $0^{+-}$ & $1_{0,2}^{--}$ & $2^{+-}$ & $3_{2,4}^{--}$ &  &  & $4$ \\ 
\hline
$3$ & $0_{1}^{++}$ & $1^{-+}$ & $2_{1,3}^{++}$ & $3^{-+}$ & $4_{3,5}^{++}$ & 
& $5$ \\ \hline
$\cdots $ & $\cdots $ & $\cdots $ & $\cdots $ & $\cdots $ & $\cdots $ & $%
\cdots $ & $\cdots $ \\ \hline
\end{tabular}
.  \label{countv}
\end{equation}
The subscripts denote the allowed values of $L$. We find $n+2$ independent
generalized form factors, as has been already shown in Ref. \cite{Ji:2000id}
(note that our $n$ differs from that in \cite{Ji:2000id}).

\subsection{Axial vector operator}

The discussion of the axial vector operator $a^{\mu }=\bar{\psi}(0)\gamma
_{5}\gamma ^{\mu }\psi (0)$ is essentially equal to that of the vector
operator (see above), only the parity and charge conjugation properties are
different. For $a^{\mu }$ we have again the two possible values $j=0,1$,
with $J^{PC}$ this time given by $1^{++}$ and $0^{-+}$. The tower of
operators 
\begin{equation}
\bar{\psi}(0)\gamma _{5}\gamma ^{\{\mu }iD^{\mu _{1}}iD^{\mu _{2}}\cdots
iD^{\mu _{n}\}}\psi (0)
\end{equation}
can be decomposed into $j=0,1,\ldots ,n+1$ angular momentum components.
Their charge parity is $C=(-)^{n}$. The parity depends on $j$ and is given
by $P=(-)^{j+1}$ (the maximal $j=n+1$ corresponding to $n+1$ spatial indices
has parity $P=(-)^{n})$, so that 
\begin{equation}
J^{PC}=j^{(-)^{j+1}(-)^{n}}=0^{(-)(-)^{n}},1^{(+)(-)^{n}},\ldots
,(n+1)^{(-)^{n}(-)^{n}}.
\end{equation}
Matching with the $\left\langle P\bar{P}\right| $-states we find 
\begin{equation}
\begin{tabular}{|l|l|l|l|l|l|l|l|}
\hline
$n\backslash J$ & $0$ & $1$ & $2$ & $3$ & $4$ & $\cdots $ & $\#$ \\ \hline
$0$ & $0_{0}^{-+}$ & $1_{1}^{++}$ &  &  &  &  & $2$ \\ \hline
$1$ & $0^{--}$ & $1_{1}^{+-}$ & $2_{2}^{--}$ &  &  &  & $2$ \\ \hline
$2$ & $0_{0}^{-+}$ & $1_{1}^{++}$ & $2_{2}^{-+}$ & $3_{3}^{++}$ &  &  & $4$
\\ \hline
$3$ & $0^{--}$ & $1_{1}^{+-}$ & $2_{2}^{--}$ & $3_{3}^{+-}$ & $4_{4}^{--}$ & 
& $4$ \\ \hline
$\cdots $ & $\cdots $ & $\cdots $ & $\cdots $ & $\cdots $ & $\cdots $ & $%
\cdots $ & $\cdots $ \\ \hline
\end{tabular}
.  \label{countav}
\end{equation}
This shows that there are $2\left\lfloor \frac{n}{2}\right\rfloor +2$
independent generalized form factors, which is in perfect agreement with the
explicit parameterization in Ref. \cite{Diehl:2003ny} (see also Eq. (\ref
{paraav}) below).

\subsection{Tensor operator}

The discussion of the (antisymmetric) tensor operator $t^{\mu \nu }=\bar{\psi%
}(0)i\sigma ^{\mu \nu }\psi (0)$ differs from that of $v^{\mu }$ and $a^{\mu
}$. The tensor operator $t^{\mu \nu }$ corresponds to the representation $%
\left( 1,0\right) \oplus \left( 0,1\right) $, therefore only $j=1$ is
possible. The charge conjugation is $C=-$. The two cases $t^{0i}$ and $t^{ik}
$, both corresponding to $j=1$, have a different parity, $P=-$ respectively $%
P=+$. This leads to the two different $J^{PC}$ components $1^{--}$ and $%
1^{+-}$. In the case of the tower of operators 
\begin{equation}
\stackunder{\lbrack \mu \nu ]}{A}\stackunder{\{\nu \mu _{1}\ldots \}}{S}\bar{%
\psi}(0)i\sigma ^{\mu \nu }iD^{\mu _{1}}iD^{\mu _{2}}\cdots iD^{\mu
_{n}}\psi (0)
\end{equation}
we first have to symmetrize and then antisymmetrize as indicated \cite
{Geyer:1999uq}. This tower corresponds to the $\left( \frac{n+2}{2},\frac{n}{%
2}\right) \oplus \left( \frac{n}{2},\frac{n+2}{2}\right) $ representation,
i.e. $j=1,\ldots ,n+1$. Again, the charge conjugation is definite and given
by $C=(-)^{n+1}$. In contrast, the parity does not only depend on $j$, but
for a given $j$ there exist in addition two possibilities corresponding to
an even or an odd number of spatial indices, as we have already seen for $%
t^{\mu \nu }$. This leads to the two sequences 
\begin{equation}
J^{PC}=j^{(-)^{j+1}(-)^{n+1}}=1^{(+)(-)^{n+1}},2^{(-)(-)^{n+1}},\ldots
,(n+1)^{(-)^{n}(-)^{n+1}}
\end{equation}
and 
\begin{equation}
J^{PC}=j^{(-)^{j}(-)^{n+1}}=1^{(-)(-)^{n+1}},2^{(+)(-)^{n+1}},\ldots
,(n+1)^{(-)^{n+1}(-)^{n+1}}.
\end{equation}
The matching with the $\left\langle P\bar{P}\right| $-states gives 
\begin{equation}
\begin{tabular}{|l|l|l|l|l|l|l|}
\hline
$n\backslash J$ & $1$ & $2$ & $3$ & $4$ & $\cdots $ & $\#$ \\ \hline
$0$ & $1_{0,2}^{--}$ &  &  &  &  & $2$ \\ \hline
$1$ & $1^{-+}$ & $2_{1,3}^{++}$ &  &  &  & $2$ \\ \hline
$2$ & $1_{0,2}^{--}$ & $2^{+-}$ & $3_{2,4}^{--}$ &  &  & $4$ \\ \hline
$3$ & $1^{-+}$ & $2_{1,3}^{++}$ & $3^{-+}$ & $4_{3,5}^{++}$ &  & $4$ \\ 
\hline
$\cdots $ & $\cdots $ & $\cdots $ & $\cdots $ & $\cdots $ & $\cdots $ & $%
\cdots $ \\ \hline
\end{tabular}
\text{ }
\begin{tabular}{|l|l|l|l|l|l|l|}
\hline
$n\backslash J$ & $1$ & $2$ & $3$ & $4$ & $\cdots $ & $\#$ \\ \hline
$0$ & $1_{1}^{+-}$ &  &  &  &  & $1$ \\ \hline
$1$ & $1_{1}^{++}$ & $2_{2}^{-+}$ &  &  &  & $2$ \\ \hline
$2$ & $1_{1}^{+-}$ & $2_{2}^{--}$ & $3_{3}^{+-}$ &  &  & $3$ \\ \hline
$3$ & $1_{1}^{++}$ & $2_{2}^{-+}$ & $3_{3}^{++}$ & $4_{4}^{-+}$ &  & $4$ \\ 
\hline
$\cdots $ & $\cdots $ & $\cdots $ & $\cdots $ & $\cdots $ & $\cdots $ & $%
\cdots $ \\ \hline
\end{tabular}
\label{countt}
\end{equation}
corresponding to a total of $2\left\lfloor \frac{n}{2}\right\rfloor
+n+3=3,4,7,8,11,12,\ldots $ generalized form factors.

\section{Parameterizations}

The decomposition of a matrix element like (\ref{ME1}) in terms of
calculable (pre-)factors given by spinor products times GFFs is not unique.
One can always rewrite the spinor products using generalized Gordon
identities \cite{Diehl:2001pm}, thereby going from one set of linear
independent GFFs to another. Once such a set of real-valued form factors for
the lowest moment has been found (under guidance of parity $\widehat{P}$ and
time reversal $\widehat{T}$), the parameterization of the higher moments
corresponding to the inclusion of covariant derivatives $iD^{\mu _{1}}\cdots
iD^{\mu _{n}}$ is essentially constructed by introducing additional factors $%
\overline{P}^{\mu _{i}}(=\left( P^{\prime \mu _{i}}+P^{\mu _{i}}\right) /2)$
and/or $\Delta ^{\mu _{i}}\Delta ^{\mu _{j}}$ $(\Delta ^{\mu _{i}}=P^{\prime
\mu _{i}}-P^{\mu _{i}})$ to the initial (lowest moment) decomposition,
respecting the $\widehat{P}$ and $\widehat{T}$ properties of the covariant
derivatives.

Before presenting the parameterization of the tower of operators involving $%
i\sigma ^{\mu \nu }$ (related to the generalized transversity), let us first
show for convenience the already known results for the vector and the axial
vector case.

\subsection{Vector operator}

The decomposition for this case has been presented in Ref. \cite{Ji:1998pc}
and is given by

\begin{eqnarray}
\left\langle P^{\prime }\right| \bar{\psi}(0)\gamma ^{\{\mu }iD^{\mu
_{1}}\cdots iD^{\mu _{n}\}}\psi (0)\left| P\right\rangle &=&\overline{U}%
(P^{\prime })\left[ \sum\Sb i=0  \\ \text{even}  \endSb ^{n}\left\{ \gamma
^{\{\mu }\Delta ^{\mu _{1}}\cdots \Delta ^{\mu _{i}}\overline{P}^{\mu
_{i+1}}\cdots \overline{P}^{\mu _{n}\}}A_{n+1,i}(\Delta ^{2})\right. \right.
\nonumber \\
&&\left. -i\frac{\Delta _{\alpha }\sigma ^{\alpha \{\mu }}{2m}\Delta ^{\mu
_{1}}\cdots \Delta ^{\mu _{i}}\overline{P}^{\mu _{i+1}}\cdots \overline{P}%
^{\mu _{n}\}}B_{n+1,i}(\Delta ^{2})\right\}  \nonumber \\
&&\left. +\left. \frac{\Delta ^{\mu }\cdots \Delta ^{\mu _{n}}}{m}%
C_{n+1,0}(\Delta ^{2})\right| _{n\text{ odd}}\right] U(P).  \label{parav}
\end{eqnarray}
On the other hand are the (moments of) spin independent GPDs expressed in
terms of polynomials in $\xi =-\left( n\cdot \Delta \right) /2$ and the GFFs

\begin{eqnarray}
H_{n+1}(\xi ,t) &\equiv &\int\limits_{-1}^{1}dxx^{n}H(x,\xi ,t)=\sum\Sb i=0 
\\ \text{even}  \endSb ^{n}\left( -2\xi \right) ^{i}A_{n+1,i}(\Delta
^{2})+\left. \left( -2\xi \right) ^{n+1}C_{n+1,0}(\Delta ^{2})\right| _{n%
\text{ odd}},\text{ }  \nonumber \\
E_{n+1}(\xi ,t) &=&\sum\Sb i=0  \\ \text{even}  \endSb ^{n}\left( -2\xi
\right) ^{i}B_{n+1,i}(\Delta ^{2})-\left. \left( -2\xi \right)
^{n+1}C_{n+1,0}(\Delta ^{2})\right| _{n\text{ odd}}.  \label{invv}
\end{eqnarray}

\subsection{Axial vector operator}

As has been observed in Ref. \cite{Diehl:2003ny}, there is no $%
C_{n+1,0}(\Delta ^{2})$-like GFF present for the axial vector, and the
parameterization shown there reads

\begin{eqnarray}
\left\langle P^{\prime }\right| \bar{\psi}(0)\gamma _{5}\gamma ^{\{\mu
}iD^{\mu _{1}}\cdots iD^{\mu _{n}\}}\psi (0)\left| P\right\rangle  &=&%
\overline{U}(P^{\prime })\sum\Sb i=0 \\ \text{even} \endSb ^{n}\left\{
\gamma _{5}\gamma ^{\{\mu }\Delta ^{\mu _{1}}\cdots \Delta ^{\mu _{i}}%
\overline{P}^{\mu _{i+1}}\cdots \overline{P}^{\mu _{n}\}}\tilde{A}%
_{n+1,i}(\Delta ^{2})\right.   \nonumber \\
&&\!\!\!\!\!\left. +\gamma _{5}\frac{\Delta ^{\{\mu }}{2m}\Delta ^{\mu
_{1}}\cdots \Delta ^{\mu _{i}}\overline{P}^{\mu _{i+1}}\cdots \overline{P}%
^{\mu _{n}\}}\tilde{B}_{n+1,i}(\Delta ^{2})\right\} U(P).  \label{paraav}
\end{eqnarray}
while the inverse relations are given by 
\begin{equation}
\widetilde{H}_{n+1}(\xi ,t)=\sum\Sb i=0 \\ \text{even} \endSb ^{n}\left(
-2\xi \right) ^{i}\widetilde{A}_{n+1,i}(\Delta ^{2}),\text{ }\widetilde{E}%
_{n+1}(\xi ,t)=\sum\Sb i=0 \\ \text{even} \endSb ^{n}\left( -2\xi \right)
^{i}\widetilde{B}_{n+1,i}(\Delta ^{2}).  \label{inva}
\end{equation}

\subsection{Tensor operator}

For the lowest moment ($n=0$) one finds three form factors, see Ref. \cite
{Diehl:2001pm},

\begin{eqnarray}
\left\langle P^{\prime }\right| \bar{\psi}(0)i\sigma ^{\mu \nu }\psi
(0)\left| P\right\rangle &=&\overline{U}(P^{\prime })\left\{ i\sigma ^{\mu
\nu }A_{T10}(\Delta ^{2})\right.  \nonumber \\
&&+\frac{\overline{P}^{[\mu }\Delta ^{\nu ]}}{m^{2}}\tilde{A}_{T10}(\Delta
^{2})  \nonumber \\
&&+\left. \frac{\gamma ^{[\mu }\Delta ^{\nu ]}}{2m}B_{T10}(\Delta
^{2})\right\} U(P).  \label{tn0}
\end{eqnarray}
Another possible structure $\varpropto \gamma ^{[\mu }\overline{P}^{\nu
]}\equiv \gamma ^{\mu }\overline{P}^{\nu }-\gamma ^{\nu }\overline{P}^{\mu }$
in (\ref{tn0}) is not allowed by time reversal symmetry, but for $n=1$ this
can be balanced with an additional factor $\Delta $, leading in agreement
with our counting to four generalized form factors (see also the detailed
discussion in the appendix of Ref. \cite{Diehl:2001pm}) ,

\begin{eqnarray}
\stackunder{\lbrack \mu \nu ]}{A}\stackunder{\{\nu \mu _{1}\}}{S}%
\left\langle P^{\prime }\right| \bar{\psi}(0)i\sigma ^{\mu \nu }iD^{\mu
_{1}}\psi (0)\left| P\right\rangle &=&\stackunder{[\mu \nu ]}{A}\stackunder{%
\{\nu \mu _{1}\}}{S}\overline{U}(P^{\prime })\left\{ i\sigma ^{\mu \nu }%
\overline{P}^{\mu _{1}}A_{T20}(\Delta ^{2})\right.  \nonumber \\
&&+\frac{\overline{P}^{[\mu }\Delta ^{\nu ]}}{m^{2}}\overline{P}^{\mu _{1}}%
\tilde{A}_{T20}(\Delta ^{2})  \nonumber \\
&&+\frac{\gamma ^{[\mu }\Delta ^{\nu ]}}{2m}\overline{P}^{\mu
_{1}}B_{T20}(\Delta ^{2})  \nonumber \\
&&+\left. \frac{\gamma ^{[\mu }\overline{P}^{\nu ]}}{m}\Delta ^{\mu _{1}}%
\tilde{B}_{T21}(\Delta ^{2})\right\} U(P).  \label{tn1}
\end{eqnarray}
Going to $n=2$, one finds by an appropriate inclusion of factors $\overline{P%
}$ and $\Delta $ a total of seven GFFs

\begin{eqnarray}
\stackunder{\lbrack \mu \nu ]}{A}\stackunder{\{\nu \mu _{1}\ldots \}}{S}%
\left\langle P^{\prime }\right| \bar{\psi}(0)i\sigma ^{\mu \nu }iD^{\mu
_{1}}iD^{\mu _{2}}\psi (0)\left| P\right\rangle &=&\stackunder{[\mu \nu ]}{A}%
\stackunder{\{\nu \mu _{1}\ldots \}}{S}\overline{U}(P^{\prime })\left\{
i\sigma ^{\mu \nu }\overline{P}^{\mu _{1}}\overline{P}^{\mu
_{2}}A_{T30}(\Delta ^{2})\right.  \nonumber \\
&&+i\sigma ^{\mu \nu }\Delta ^{\mu _{1}}\Delta ^{\mu _{2}}A_{T32}(\Delta
^{2})  \nonumber \\
&&+\frac{\overline{P}^{[\mu }\Delta ^{\nu ]}}{m^{2}}\overline{P}^{\mu _{1}}%
\overline{P}^{\mu _{2}}\tilde{A}_{T30}(\Delta ^{2})  \nonumber \\
&&+\frac{\overline{P}^{[\mu }\Delta ^{\nu ]}}{m^{2}}\Delta ^{\mu _{1}}\Delta
^{\mu _{2}}\tilde{A}_{T32}(\Delta ^{2})  \nonumber \\
&&+\frac{\gamma ^{[\mu }\Delta ^{\nu ]}}{2m}\overline{P}^{\mu _{1}}\overline{%
P}^{\mu _{2}}B_{T30}(\Delta ^{2})  \nonumber \\
&&+\frac{\gamma ^{[\mu }\Delta ^{\nu ]}}{2m}\Delta ^{\mu _{1}}\Delta ^{\mu
_{2}}B_{T32}(\Delta ^{2})  \nonumber \\
&&+\left. \frac{\gamma ^{[\mu }\overline{P}^{\nu ]}}{m}\Delta ^{\mu _{1}}%
\overline{P}^{\mu _{2}}\tilde{B}_{T31}(\Delta ^{2})\right\} U(P).
\label{tn2}
\end{eqnarray}
Continuing this chain of reasoning, we need as shown above a total of $%
2\left\lfloor \frac{n}{2}\right\rfloor +n+3$ generalized form factors for
the parameterization of the n$\emph{th}$ moment of the tensor operator

\begin{eqnarray}
&&\stackunder{\lbrack \mu \nu ]}{A}\stackunder{\{\nu \mu _{1}\ldots \}}{S}%
\left\langle P^{\prime }\right| \bar{\psi}(0)i\sigma ^{\mu \nu }iD^{\mu
_{1}}\cdots iD^{\mu _{n}}\psi (0)\left| P\right\rangle   \nonumber \\
&=&\stackunder{[\mu \nu ]}{A}\stackunder{\{\nu \mu _{1}\ldots \}}{S}%
\overline{U}(P^{\prime })\left[ \sum\Sb i=0 \\ \text{even} \endSb %
^{n}\left\{ i\sigma ^{\mu \nu }\Delta ^{\mu _{1}}\cdots \Delta ^{\mu _{i}}%
\overline{P}^{\mu _{i+1}}\cdots \overline{P}^{\mu _{n}}A_{Tn+1,i}(\Delta
^{2})\right. \right.   \nonumber \\
&&+\frac{\overline{P}^{[\mu }\Delta ^{\nu ]}}{m^{2}}\Delta ^{\mu _{1}}\cdots
\Delta ^{\mu _{i}}\overline{P}^{\mu _{i+1}}\cdots \overline{P}^{\mu _{n}}%
\tilde{A}_{Tn+1,i}(\Delta ^{2})  \nonumber \\
&&+\left. \frac{\gamma ^{[\mu }\Delta ^{\nu ]}}{2m}\Delta ^{\mu _{1}}\cdots
\Delta ^{\mu _{i}}\overline{P}^{\mu _{i+1}}\cdots \overline{P}^{\mu
_{n}}B_{Tn+1,i}(\Delta ^{2})\right\}   \nonumber \\
&&+\left. \sum\Sb i=0 \\ \text{odd} \endSb ^{n}\frac{\gamma ^{[\mu }%
\overline{P}^{\nu ]}}{m}\Delta ^{\mu _{1}}\cdots \Delta ^{\mu _{i}}\overline{%
P}^{\mu _{i+1}}\cdots \overline{P}^{\mu _{n}}\tilde{B}_{Tn+1i}(\Delta
^{2})\right] U(P).  \label{parat}
\end{eqnarray}
The polynomial relations between the generalized transversity plus the other
corresponding GPDs and the GFFs are 
\begin{eqnarray}
H_{T,n+1}(\xi ,t) &=&\sum\Sb i=0 \\ \text{even} \endSb ^{n}\left( -2\xi
\right) ^{i}A_{Tn+1,i}(\Delta ^{2}),\text{ }\widetilde{H}_{T,n+1}(x,\xi
,t)=\sum\Sb i=0 \\ \text{even} \endSb ^{n}\left( -2\xi \right) ^{i}%
\widetilde{A}_{Tn+1,i}(\Delta ^{2}),  \nonumber \\
E_{T,n+1}(\xi ,t) &=&\sum\Sb i=0 \\ \text{even} \endSb ^{n}\left( -2\xi
\right) ^{i}B_{Tn+1,i}(\Delta ^{2}),\text{ }\widetilde{E}_{T,n+1}(x,\xi
,t)=\sum\Sb i=0 \\ \text{odd} \endSb ^{n}\left( -2\xi \right) ^{i}\widetilde{%
B}_{Tn+1,i}(\Delta ^{2}),  \label{invt}
\end{eqnarray}
showing explicitely that $\widetilde{E}_{T}(x,\xi ,t)$ is the only twist-2
GPD being antisymmetric in $\xi $ \cite{Diehl:2001pm}.

\section{Summary}

Based on the method presented in Ref. \cite{Ji:2000id} we have counted the
number of independent generalized form factors parameterizing the towers of
axial vector and tensor operators, see tables (\ref{countav},\ref{countt}).
This gave us an independent check on the actual decomposition of the tensor
operator, which is presented in Eq. (\ref{parat}). Taking together these
results with the corresponding representations of the tensor operator on a
space-time-lattice \cite{Meinulf2004} will probably allow for a first
determination of the lowest moments of the generalized transversity in
lattice QCD.

\subsection*{Acknowledgments}

The author would like to thank D. Boer, M. Diehl, M. G\"{o}ckeler and A. Sch%
\"{a}fer for helpful discussions.


\begin{thebibliography}{99}
\bibitem{Ji:1996ek}  X.~D.~Ji, 
%``Gauge invariant decomposition of nucleon spin,''
Phys.\ Rev.\ Lett.\ \textbf{78} (1997) 610 [arXiv:hep-ph/9603249]. 
%%CITATION = HEP-PH 9603249;%%

\bibitem{Diehl:2001pm}  M.~Diehl, 
%``Generalized parton distributions with helicity flip,''
Eur.\ Phys.\ J.\ C \textbf{19} (2001) 485 [arXiv:hep-ph/0101335]. 
%%CITATION = HEP-PH 0101335;%%

%\cite{:2003is}

\bibitem{Hoodbhoy:1998vm}  P.~Hoodbhoy and X.~D.~Ji, 
%``Helicity-flip off-forward parton distributions of the nucleon,''
Phys.\ Rev.\ D \textbf{58} (1998) 054006 [arXiv:hep-ph/9801369]. 
%%CITATION = HEP-PH 9801369;%%

\bibitem{lattice1}  P.~H\"{a}gler, J.~W.~Negele, D.~B.~Renner, W.~Schroers,
T.~Lippert and K.~Schilling [LHPC Collaboration], [arXiv:hep-lat/0312014].

\bibitem{lattice2}  P.~H\"{a}gler, J.~W.~Negele, D.~B.~Renner, W.~Schroers,
T.~Lippert and K.~Schilling [LHPC Collaboration], Phys.\ Rev.\ D \textbf{68}
(2003) 034505 [arXiv:hep-lat/0304018].

\bibitem{lattice3}  M.~G\"{o}ckeler, R.~Horsley, D.~Pleiter, P.~E.~L.~Rakow,
A.~Schafer, G.~Schierholz and W.~Schroers [QCDSF Collaboration], 
%``Generalized parton distributions from lattice QCD,''
Phys.\ Rev.\ Lett.\ \textbf{92} (2004) 042002 [arXiv:hep-ph/0304249]. 
%%CITATION = HEP-PH 0304249;%%

\bibitem{Ji:2000id}  X.~D.~Ji and R.~F.~Lebed, 
%``Counting form factors of twist-two operators,''
Phys.\ Rev.\ D \textbf{63} (2001) 076005 [arXiv:hep-ph/0012160]. 
%%CITATION = HEP-PH 0012160;%%

%\cite{Ji:1996ek}

\bibitem{Geyer:1999uq}  B.~Geyer, M.~Lazar and D.~Robaschik, 
%``Decomposition of nonlocal light-cone operators into harmonic operators  of
%definite twist,''
Nucl.\ Phys.\ B \textbf{559} (1999) 339 [arXiv:hep-th/9901090]. 
%%CITATION = HEP-TH 9901090;%%

%\cite{Weinberg:mt}

\bibitem{Weinberg:mt}  S.~Weinberg, ``The Quantum Theory Of Fields. Vol. 1:
Foundations,'',Cambridge, UK: Univ. Pr. (1995). 
%\href{http://www.slac.stanford.edu/spires/find/hep/www?irn=3355144}{SPIRES entry}

%\cite{Hoodbhoy:1998yb}

\bibitem{Diehl:2003ny}  M.~Diehl, %``Generalized parton distributions,''
Phys.\ Rept.\ \textbf{388} (2003) 41 [arXiv:hep-ph/0307382]. 
%%CITATION = HEP-PH 0307382;%%

\bibitem{Ji:1998pc}  X.~D.~Ji, %``Off-forward parton distributions,''
J.\ Phys.\ G \textbf{24} (1998) 1181 [arXiv:hep-ph/9807358]. 
%%CITATION = HEP-PH 9807358;%%

%\cite{Hoodbhoy:1998vm}

\bibitem{Meinulf2004}  M. G\"{o}ckeler, private communication.
\end{thebibliography}
\end{document}